\documentclass[12pt,preprint]{aastex}

\newcommand{\dechms}[4]{$#1^{\rm h}#2^{\rm m}#3\mbox{$^{\rm s}\mskip-7.6mu.\,$}#4$}

\newcommand{\decdms}[4]{$+#1^{\circ}#2'#3\mbox{$''\mskip-7.6mu.\,$}#4$}

\slugcomment{To appear in AJ}

\shorttitle{SMA Submillimeter Observations of HL Tau}
\shortauthors{Lumbreras \& Zapata}

\begin{document}


\title{SMA Submillimeter Observations of HL Tau: Revealing a compact molecular outflow}


\author{Alba M. Lumbreras,\altaffilmark{1,2} and Luis A. Zapata\altaffilmark{1}}

\altaffiltext{1}{Centro de Radioastronom\'\i{}a y Astrof\'\i{}sica,
  UNAM, M\'exico.}
\altaffiltext{2}{Benem\'erita Universidad Aut\'onoma de Puebla, Puebla, M\'exico}

\begin{abstract}
We present archival high angular resolution ($\sim$ 2$''$)  
$^{12}$CO(3-2) line and continuum submillimeter observations 
of the young stellar object HL Tau made with the Submillimeter Array (SMA). 
The $^{12}$CO(3-2) line observations reveal the presence 
of a compact and wide opening angle bipolar outflow with a 
northeast and southwest orientation (P.A. = 50$^\circ$), and that is
associated with the optical and infrared jet emanating from HL Tau with 
a similar orientation.  On the other hand, the 850 $\micron$ continuum emission 
observations exhibit a strong and compact source in the position of HL Tau that 
has a spatial size of $\sim$ 200 $\times$ 70 AU with a P.A. $=$ 145$^\circ$, 
and a dust mass of around 0.1 M$_\odot$. These physical parameters are in agreement with values obtained 
recently from millimeter observations. This submillimeter source is therefore related with the disk surrounding HL Tau.
\end{abstract}

\keywords{ISM: Jets and outflows; Stars: Pre-main sequence; Stars: Mass loss}

\section{Introduction}

HL Tau is a young star of solar-type located in the Taurus molecular cloud, a close-by 
star-forming region at a distance of about 140 pc \citep{tor2007,red2004}. This young star has been
extensively studied at many wavelengths, some of these include radio, millimeter, infrared, and optical 
\citep{mun1996, wil1996, ang2007, tak2007, kri2008,car2009, kow2011}. These observations have revealed that 
HL Tau possesses yet a well defined disk-outflow system.  The circumstellar disk has a 
mass of 0.13 M$_\odot$, a characteristic radius of 80 AU, and a position angle P.A. = 136$^\circ$ \citep{kow2011}. 
On the other hand, the infrared and optical observations show the presence of a    
collimated jet emanating from HL Tau with an orientation northeast and southwest \citep[P.A. = 50$^\circ$;][]{kri2008}. 
At very small scales ($\sim$ 50 AU) a radio thermal jet has also been reported to be 
associated with HL Tau with a similar orientation 
to that shown by the optical and infrared jet \citep{rod1994, car2009}.  \citet{rot2007} confirmed  
that HL Tau has a large dusty envelope discovered through modeling to a spectral 
energy distribution over different frequencies. This envelope was reported for the first time 
by \citet{cab1996}. 

At large scales ($\sim$ 2000 AU), using single dish millimeter telescopes, \citet{mon1996, cab1996} found 
a very extended northeast and southwest molecular outflow emanating from HL Tau. This flow was reported 
as an anisotropic, mostly redshifted that is emanating from within a flattened molecular remnant envelope. 
\citet{cab1996} also showed Plateau de Bure interferometer (PdBI) maps that revealed small-scale structures nested 
in the putative extended molecular outflow.     
    
Here, we present submillimeter observations that reveal a very well defined molecular compact   
and wide opening angle bipolar outflow with a northeast and southwest orientation that is
likely associated with the large-scale monopolar molecular outflow reported by \citet{mon1996, cab1996}. 
The 850 $\micron$ continuum images reveal that the dust emission at this wavelength 
is mainly associated with its circumstellar disk.

\begin{figure*}[!t]
  \centering
  \includegraphics[angle=0,scale=0.48]{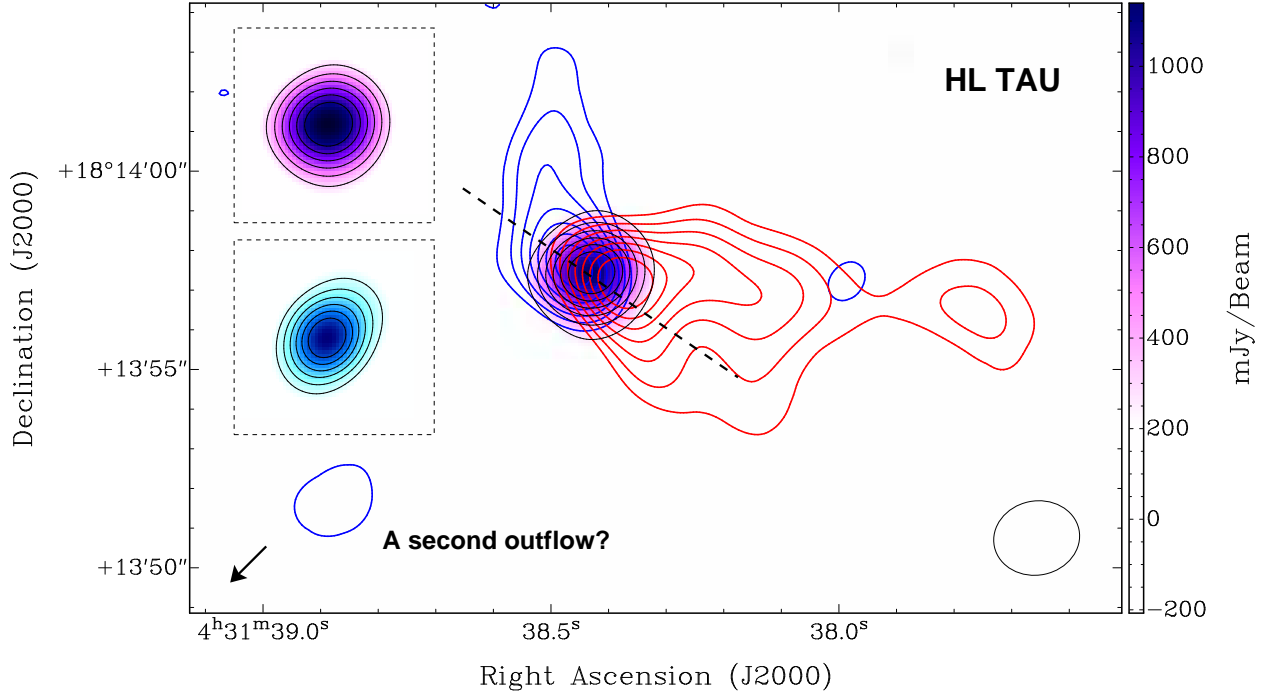}
  \caption{ \scriptsize High-velocity SMA integrated intensity (moment 0) contour (blue and red) map of the $^{12}$CO(3-2) 
                thermal emission, and the continuum emission at 850 $\micron$ 
                (yellow scale and black contours)  from HL Tau.  The blue contours represent blueshifted gas, while
                the red contours represent redshifted gas.
                The blue and red contours range from, 30\% to 90\% of the peak emission, in steps of 10\%. 
                The emission peaks for the blueshifted and redshifted emission are 6.5 and 19.1 Jy Beam$^{-1}$
                 km s$^{-1}$, respectively. The black contours range from 35\% to 90\% of the peak emission, 
                 in steps of 10\%. The peak of the continuum emission is 1.0 Jy beam$^{-1}$. 
                 The synthesized beam of the continuum image is shown in the lower right corner. 
                 The dashed black line with a P.A. of 50$^\circ$
                 traces the position where the position-velocity diagram shown in Figure 3 was computed. 
                 The black arrow marks the orientation of a possible second outflow emanating from HL Tau. 
                 The CO synthesized beam is very similar to that of the continuum image, see text.
                 We show in the upper left corner the continuum source (purple scale) and the resulting continuum
                 source (blue scale) after restoring it with a beam size slightly smaller of  $1\rlap.{''}7$ $\times$ $1\rlap.{''}7$ 
                 with a P.A. = $+$85$^\circ$. For the blue scale image, the contouring is the same as above but with 
                 the peak at 0.6 Jy Beam$^{-1}$. The negative contours are not shown in this image.}
  \label{fig:fig1}
\end{figure*}

\section{Observations}

The observations were obtained from the SMA archive, and were collected on October 2005, 
when the array was in its compact configuration.  The 21 independent baselines in the compact 
configuration ranged in projected length from 13 to 79 k$\lambda$.  The phase reference center 
for the observations was at $\alpha_{J2000.0}$ = \dechms{04}{31}{38}{41}, $\delta_{J2000.0}$ = 
\decdms{18}{13}{57}{79}. Two frequency bands, centered at 336.5 GHz (Lower Sideband) and 
346.5 GHz (Upper Sideband) were observed simultaneously. The primary beam of the SMA at 
345 GHz has a FWHM $\sim 30''$.  The submillimeter emission arising from HL Tau falls very 
well inside of the FWHM. 

The SMA digital correlator was configured in 24 spectral windows (``chunks'') of 104 MHz and
128 channels each. This provides a spectral resolution of 0.815 MHz ($\sim$ 0.7 km s$^{-1}$) 
per channel.  The system temperatures (T$_{DSB}$) varied in a range of 200 to 450 K, indicating 
good weather conditions.  Observations of Uranus provided the absolute scale for the flux calibration.
The gain calibrator was the quasar 3C 111, while Uranus was also used for bandpass calibration. 
The uncertainty in the flux scale is estimated to be between 15 and 20$\%$, based on the SMA monitoring of 
quasars.  Further technical descriptions of the SMA can be found in \citet{Hoetal2004}.

The data were calibrated using the IDL superset MIR, originally developed for the Owens Valley 
Radio Observatory \citep[OVRO,][]{Scovilleetal1993} and adapted for the SMA.\footnote{The 
MIR-IDL cookbook by C. Qi can be found at http://cfa-www.harvard.edu/$\sim$cqi/mircook.html.} 
The calibrated data were then imaged and analyzed in the standard manner using the MIRIAD and
KARMA \citep{goo96,sau1995} softwares. A  850 $\micron$ continuum image was obtained by 
averaging line-free channels in the upper sideband with a total bandwidth of 2 GHz.   
For the line emission, the continuum was also removed. The $^{12}$CO(3-2) was the only spectral line detected.
For the continuum emission, we set the {\emph ROBUST} parameter of the task {\emph INVERT} to -2 to obtain
a slightly higher angular resolution allowing us to obtain a better fitting for the size of HL Tau, 
while for the line emission we set this to $+$2 in order to obtain a better sensitivity sacrificing angular resolution.

The resulting r.m.s.\ noise for the 
continuum image was about 20 mJy beam$^{-1}$ at an angular resolution of $2\rlap.{''}07$ 
$\times$ $1\rlap.{''}88$ with a P.A. = $+$85.2$^\circ$. The r.m.s.\ noise in each channel of the 
spectral line data was about 200 mJy beam$^{-1}$ at an angular resolution of $2\rlap.{''}38$ 
$\times$ $2\rlap.{''}19$ with a P.A. = $+$14.5$^\circ$.




\begin{figure*}[!t]
  \centering
  \includegraphics[angle=-90, scale=0.57]{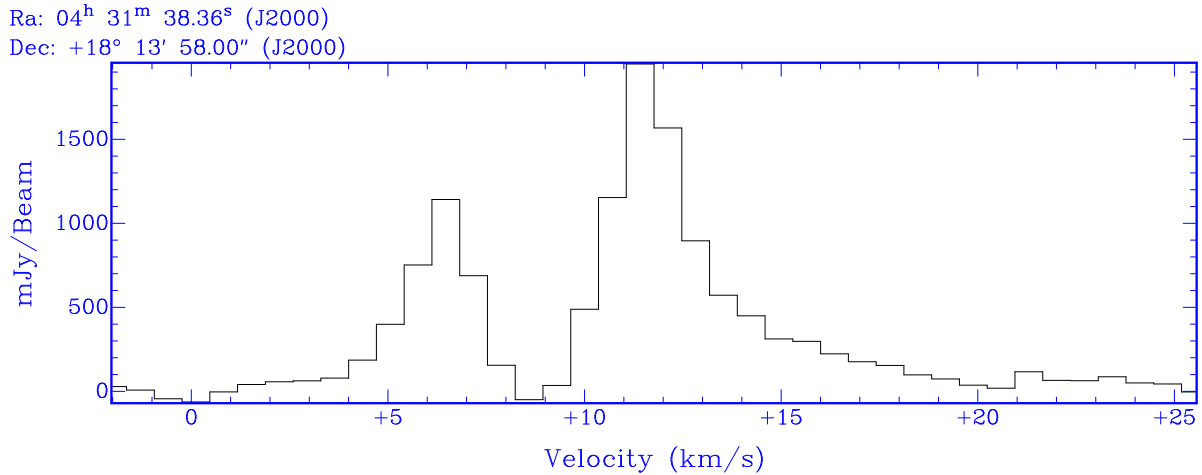}
  \caption{ \scriptsize Average spectra of the $^{12}$CO(3-2) line towards HL Tau. On top of the image
   is shown the central position of the box, where the spectra was obtained. The spectra was 
   obtained from an area delimited by a $5" \times 5"$ square. The velocity here is LSR.}
  \label{fig:fig2}
\end{figure*}

\section{Results}
\label{sec:resu}

\subsection{Continuum emission}

In Figure 1, we show the resulting 850 $\micron$ continuum image from HL Tau made with the SMA. 
We only detected a single source that is associated with HL Tau. This source has a position in the sky
of $\alpha_{J2000.0}$ = \dechms{04}{31}{38}{42}, $\delta_{J2000.0}$ = 
\decdms{18}{13}{57}{39}, with a positional error of less than $1\rlap.{''}$0. This position coincides 
very well within the uncertainty with the position obtained by \citet{mun1996} at 2.7 mm with BIMA,
discarding any proper motion of HL Tau in this interval of time.

We measured a flux density and peak intensity for this compact source  
of 1.3$\pm$0.3 Jy and 1.0$\pm$0.2 Jy Beam$^{-1}$, respectively. These flux values are in  good
agreement with the SED for HL Tau obtained in \citet{kow2011}. In Figure 1, HL Tau looks something rounded, 
however, this is resolved as an elongated source using the task of MIRIAD {\it imfit} and 
{\it jmfit} of AIPS. These two tasks use elliptical gaussian fittings to estimate 
the source parameters.
The deconvolved size that we get is that one for an elongated source with a northwest-southeast orientation    
 (approximately $1\rlap.{''}5$ $\times$ $0\rlap.{''}5$  with a P.A. = $+$145$^\circ$).
This angular size corresponds to $\sim$ 200 $\times$ 70 AU 
with a P.A. $=$ 145$^\circ$ at the distance of the Taurus molecular cloud.
These size values also are in good agreement with those values already obtained by \citet{kow2011} using
CARMA 1.3 and 2.7 mm observations.
Given the high signal-to-noise ratio obtained from our observations, we tried to verify 
the results obtained by the tasks {\it imfit} and {\it jmfit}, and we then restored our original image 
with a slightly smaller beam (see Figure 1) revealing the elongated morphology obtained by the tasks.
 
Assuming optically thin isothermal dust emission, a gas-to-dust ratio of 
100, a dust temperature of 30 K, a dust mass opacity $\kappa_{850 \micron}$ = 1.5 cm$^2$ g$^{-1}$, 
and an emissivity index $\beta = 1.0$, \citep[see][]{mun1996,kow2011}, 
we estimate roughly a total mass for the source of 0.1$\pm$0.02 M$_\odot$. 
The uncertainty here is only the error in the flux measurement of HL Tau.
However, we noted that the temperature for the disk used here could be higher as 
shown in the model presented in Kwon et al. (2011).

This combination of properties suggest that the emission seen at  850 $\micron$ is dominated by the
accretion disk, although a very small contribution from the inner envelope might also be present.

\subsection{$^{12}$CO(3-2) line emission}

In Figure 2, we show the average spectrum of the $^{12}$CO(3-2) line towards HL Tau.
The box length of the averaging spectra is about 5 arcseconds centered in the position of HL Tau.
The spectrum has two strong peaks,  the first peak of the spectrum appears to be at
a velocity slightly smaller than 6.5 km/s, and the other one about 11.5 km s$^{-1}$.
The systemic velocity of the cloud is $\sim$ 6.5 km/s (Monin et al. 1996 and Cabrit et al. 1996). 
The blueshifted peak is probably arising from the outflow and disk as revealed by Cabrit et al. (1996). 
The blueshifted peak is fainter than the redshifted one. There is one dip feature about 8.5 km s$^{-1}$, maybe caused 
for the missing flux at velocities close to the systemic. In addition, there are some faint high velocity features associated with the outflow emanating from HL Tau.    

In Figure 1, additionally to the 850 $\micron$ continuum image, we have overlaid the
integrated intensity (moment 0) high-velocity map of the $^{12}$CO(3-2) thermal emission.  In the image
the blue and red colors correspond to the blueshifted and redshifted gas emission, respectively. To construct
the moment 0 high-velocity maps, we have integrated LSR velocities ranging from $+$1.88 km s$^{-1}$ to $+$5.1 km s$^{-1}$
for the blue shifted one, and LSR velocities ranging from $+$13.24 km s$^{-1}$ to 
$+$25.53 km s$^{-1}$ for the redshifted.  The emission at ambient velocities ($+$6 to $+$11 km s$^{-1}$) 
was clearly extended and poorly sampled with the SMA, and was suppressed in this moment zero map.
This extended emission is arising mainly from the molecular cloud, and 
thus the systemic velocity of the cloud and HL Tau should be placed within this range.

This image reveals a compact bipolar molecular outflow emanating 
from HL Tau. The bipolar outflow has a northeast and southwest axis with a position angle of about    
50$^\circ$. We thus noted that this bipolar outflow is the molecular counterpart of the optical and 
infrared jet that emanates from HL Tau with a similar position angle \citep{ang2007, tak2007, kri2008}. 
However, contrary of being collimated as the optical and infrared jet, and other bipolar molecular outflows {\it e.g.} HH 797 or 
HH 625 \citep{zap2005,pec2012}, the molecular counterpart is a wide opening angle outflow and is only present
very close to HL Tau. These physical properties have also been observed in other outflows, for example, in
the HH 30, and are attributed to the entrainment by the optical jet \citep{pet2006}.

We note that both the blue and red northwest walls appear to be significantly 
more defined than the southeast ones. A possibility is that this effect could be created 
if very dense molecular material is present in the west part of the cloud. This condition
will help to a better entrainment of the molecular gas material. 

\begin{figure}[!t]
  \includegraphics[angle=0,width=0.9\columnwidth]{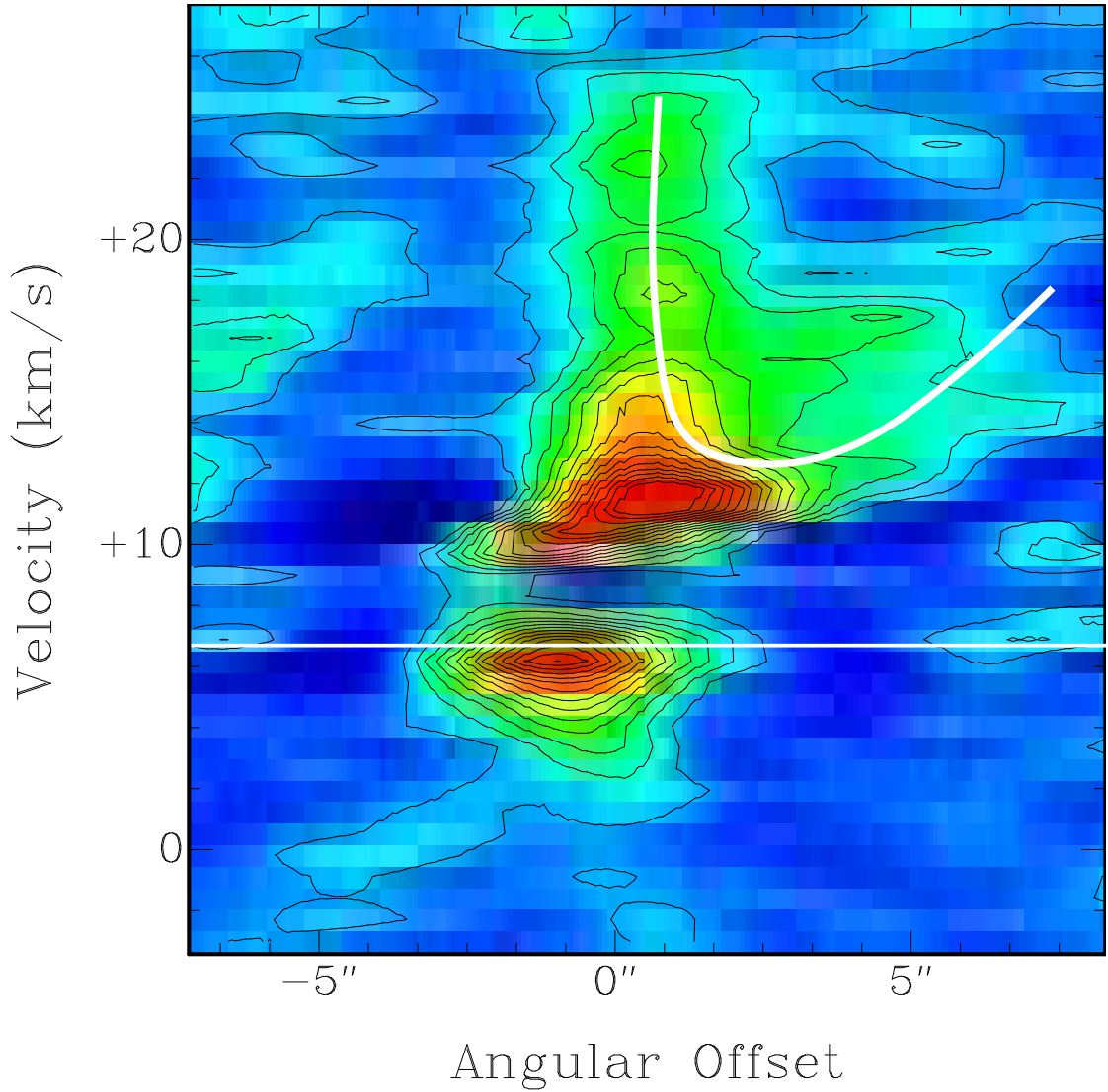}
  \caption{\scriptsize  Position-velocity diagram of the CO outflow computed at a position angle of $+$50$^\circ$. 
               The contours range from 20\% to 90\% of the peak emission, in steps of 5\%. 
               The peak of the line emission is 6.5 Jy beam$^{-1}$. The systemic LSR radial velocity 
               of the ambient molecular cloud is about $+$6.5 km s$^{-1}$ and is represented with a straight line. 
               The synthesized beam is 2.38$''$ $\times$ 2.19$''$ with a P.A. of $+$14.5$^\circ$, and the spectral resolution
               is $\sim$ 0.7 km s$^{-1}$. The white curved line illustrates the wide-angle feature observed in its redshifted 
               side. The negative contours are not shown in this image. }
  \label{fig:fig3}
\end{figure}

The spatial structure of the moderate/low velocity gas of the bipolar outflow as a function of the
radial velocities (position-velocity diagram) along the mayor axis is shown in Figure 3. 
This image shows how the redshifted component covers a much wider range of velocities than 
the blueshifted side. The redshifted lobe displays a high velocity component with a low-velocity
component extending far from the position of the star.  On the other hand, 
the blue side of the outflow displays only a limited range of velocities close to the exciting source.  

Based on Figures presented by \citet{arce2007,lee2000} that show the molecular outflow 
properties predicted by different models, 
and in the morphology and the PV diagram obtained here from the outflow (Figures 1 and 3), 
{we suggest that this outflow is reminiscent of a wide-angle outflow. 
However, we noted that there is also a collimated jet emanating simultaneously 
from HL Tau mapped in the optical and infrared wavelengths  \citep{kri2008}. This is therefore one of the outflows that
seems to present simultaneously a jet and a wide-angle outflow \citep[see][]{arce2007}.
However, many more sensitive high-angular resolution observations are needed to resolve better the putative
wide-angle outflow and the jet.

Assuming that we are in local thermodynamic equilibrium (LTE), that the $^{12}$CO(3-2) molecular emission 
is optically thin, a fractional abundance between the carbon monoxide and the molecular hydrogen of 10$^4$,
a distance of 140 pc, an excitation temperature T$_{ex}$ = 50 K, we estimated a mass for the outflow energized
by HL Tau of 2.5 $\times$ 10$^{-3}$ M$_\odot$. This value for the outflow mass is expected for a flow energized 
by a young low-mass protostar, see \citet{wu2004}. The uncertainty in the outflow mass is on the order of a factor 
of two, mainly due to the ambiguity of the excitation temperature. We also estimate  
a kinematical energy of  
3 $\times$ 10$^{42}$ ergs, 
an outflow momentum of 0.03 M$_\odot$ km s$^{-1}$.   
These estimates do not take into consideration the inclination
 of the outflow with respect to the line-of-sight and the amplitude calibration. 

We found some faint molecular emission toward the southeast of HL Tau (about 10$''$),
where is located the optical object HH 153,  see Figure 1.  The emission revealed in Figure 1
seems to be marginal, however this emission is present in a few velocity channels ($-$0.88 to $+$1.24 km s$^{-1}$) 
making more reliable our detection.  The HH 153 is a traverse outflow that appears 
to emanate from the HL Tau vicinity \citep{kri2008}. However, we note that the measured proper 
motions of HH 153 (knots Halpha-B1, B2, B3, C1, C2; Anglada et al. 2007) point away from HL 
Tau supporting also an origin in this source. Future sensitive molecular observations of this region 
may help in revealing its energizing object.    

\section{Conclusions}
\label{sec:resu}

We have observed in the submillimeter regime the young star HL Tau using 
the Submillimeter Array.  Our conclusions are as follow:

\begin{itemize} 
\item The $^{12}$CO(3-2) line observations reveal the presence 
of a compact and wide opening angle bipolar outflow with a 
northeast and southwest orientation (P.A. = 50$^\circ$), and that is
associated with the optical and infrared jet emanating from HL Tau with 
a similar orientation. The compact outflow has a mass of about 2.5 $\times$ 10$^{-3}$ M$_\odot$.

\item We found that in the redshifted side of the outflow there is evidence of
a wide-angle outflow that probably is associated with an optical jet. 
This outflow is likely associated with the large-scale monopolar molecular outflow 
reported by \citet{mon1996, cab1996}.

\item The 850 $\micron$ continuum emission 
observations show a strong and compact source in the position of HL Tau itself  that 
has a deconvolved size of $\sim$ 200 $\times$ 70 AU with a P.A. $=$ 145$^\circ$ and a mass of 0.1 M$_\odot$.

\end{itemize} 
 
 \acknowledgments
 
L.A.Z. acknowledge the financial support from DGAPA, UNAM, and CONACyT, M\'exico.
We would like to thank to the anonymous referee for the comments, which definitively 
help to improve our manuscript.



{\it Facilities:} \facility{SMA}.



\end{document}